\documentclass[a4paper,11pt]{article}
\pdfoutput=1 

\usepackage{jcappub} 

\usepackage[T1]{fontenc} 
\usepackage{multirow}
\usepackage{natbib}
\usepackage{aas_macros}
\usepackage{amsmath}
\usepackage{color}
\usepackage{enumerate}
\usepackage{multirow}
\usepackage{bbm}
\usepackage{hyperref}
\usepackage{tabularx}
\usepackage{xfrac}
\usepackage{arydshln}

\usepackage{caption}
\usepackage{newfloat}
\usepackage{algorithm}
\usepackage{algorithmic}

\DeclareFloatingEnvironment[name={List}]{enumcnt}

\definecolor{internationalkleinblue}{rgb}{0.0, 0.18, 0.65}

\hypersetup{urlcolor=internationalkleinblue, linkcolor=internationalkleinblue, citecolor=internationalkleinblue}

\newcommand\Tstrut{\rule{0pt}{3ex}}   

\title{Integrated cosmological probes: Concordance quantified}

\author{Andrina Nicola,}
\author{Adam Amara,}
\author{Alexandre Refregier}

\affiliation{Department of Physics, ETH Z\"urich, Wolfgang-Pauli-Strasse 27, CH-8093 Z\"urich, Switzerland}

\emailAdd{andrina.nicola@phys.ethz.ch}

\abstract{Assessing the consistency of parameter constraints derived from different cosmological probes is an important way to test the validity of the underlying cosmological model. In an earlier work \cite{Nicola:2017}, we computed constraints on cosmological parameters for $\Lambda$CDM from an integrated analysis of CMB temperature anisotropies and CMB lensing from Planck, galaxy clustering and weak lensing from SDSS, weak lensing from DES SV as well as Type Ia supernovae and Hubble parameter measurements. In this work, we extend this analysis and quantify the concordance between the derived constraints and those derived by the Planck Collaboration as well as WMAP9, SPT and ACT. As a measure for consistency, we use the Surprise statistic \cite{Seehars:2014}, which is based on the relative entropy. In the framework of a flat $\Lambda$CDM cosmological model, we find all data sets to be consistent with one another at a level of less than 1$\sigma$. We highlight that the relative entropy is sensitive to inconsistencies in the models that are used in different parts of the analysis. In particular, inconsistent assumptions for the neutrino mass break its invariance on the parameter choice. When consistent model assumptions are used, the data sets considered in this work all agree with each other and $\Lambda$CDM, without evidence for tensions.}

\begin{document}
\maketitle
\flushbottom

\section{Introduction}\label{sec:intro}

In recent years, many large cosmological surveys have been conducted and many more are currently in the planning \cite{WMAP, Planck, DES, DESI, Euclid, LSST, WFIRST}. These surveys have allowed us to further constrain our cosmological model, leading to the establishment of $\Lambda$CDM (see e. g. Ref.~\cite{Planck-Collaboration:2016ae} and references therein). Joint cosmological probe analyses are an important way to test some of the key components of this model such as Dark Matter and Dark Energy. In addition, it may be illuminating to assess if different cosmological probes are consistent with one another within the same cosmological model. Inconsistencies might be a sign for tensions within the cosmological model assumed and systematics in the measurements considered.

In recent work \cite{Nicola:2016, Nicola:2017}, we computed constraints on cosmological parameters for a $\Lambda$CDM cosmological model from an integrated analysis (IA) of Cosmic Microwave Background (CMB) temperature anisotropies and CMB lensing maps from the Planck mission \cite{Planck-Collaboration:2016ad}, galaxy clustering from the eighth data release of the Sloan Digital Sky Survey (SDSS DR8) \cite{Aihara:2011}, weak lensing from both SDSS Stripe 82 \cite{Annis:2014} and the Dark Energy Survey (DES) Science Verification (SV) data \cite{Jarvis:2016} as well as Type Ia supernovae (SNe Ia) data from the joint light curve analysis (JLA) \cite{Betoule:2014}, and constraints on the Hubble parameter from the Hubble Space Telescope (HST) \cite{Riess:2011, Efstathiou:2014}\footnote{We note that the latter two analyses are not completely independent since they both contain low redshift SNe Ia from CfAIII \cite{Hicken:2009}. In this work however, we do not take into account this correlation and treat the two analyses as independent. We thank our referee for pointing this out.}. Comparing these results to those obtained by the Planck Collaboration \cite{Planck-Collaboration:2016ae} and those from WMAP9 \cite{Hinshaw:2013}, we found a good agreement between the IA constraints and those from WMAP9 while the marginalised 2D contours suggested hints of tensions with the results from Planck 2015. Impressions gained through 2D contours however, can be misleading as shown by e.g. Ref.~\cite{Seehars:2016}. In this work, we therefore extend our previous analysis \cite{Nicola:2016, Nicola:2017} by quantifying the concordance between the results derived from these different data sets. Since for consistency with Ref.~\cite{Planck-Collaboration:2016ae}, the IA does not include the latest Hubble parameter measurement presented in Ref.~\cite{Riess:2016}, we further investigate the impact of our choice of Hubble parameter measurement on the concordance between the data sets considered.

Several different concordance measures have been proposed in the literature (see e.g. \cite{Hobson:2002, Kunz:2006, Raveri:2016, Charnock:2017, Lin:2017}). In the present work, we employ the Surprise statistic, which was introduced in Ref.~\cite{Seehars:2014} and is based on the Kullback-Leibler divergence \cite{Kullback:1951} or relative entropy. 
The Surprise has been applied to data from the CMB, Large Scale Structure (LSS) as well as background probes \cite{Seehars:2014, Seehars:2016, Grandis:2016, Grandis:2016aa, Zhao:2017} and has been shown to be a robust measure for concordance in cosmology. In the framework of a $\Lambda$CDM cosmological model we use the Surprise to compare the constraints obtained from the integrated analysis in Ref.~\cite{Nicola:2017} to several other data sets under the assumption of Gaussianity of all posterior distributions. We consider the constraints derived by the Planck Collaboration \cite{Planck-Collaboration:2016ae}, those derived from the combination of WMAP9 data \cite{Hinshaw:2013} with data from the Atacama Cosmology Telescope (ACT) \cite{Fowler:2010, Das:2011} and the South Pole Telescope (SPT) \cite{Keisler:2011, Reichardt:2012} as well as the recent update to the WMAP9+SPT+ACT constraints from Ref.~\cite{Calabrese:2017}. 

This paper is organised as follows. We review the Surprise statistic in Section \ref{sec:methods} while Section \ref{sec:data} describes the data sets used in this work. Our results are presented in Section \ref{sec:results} and we conclude in Section \ref{sec:conclusions}. Robustness tests are deferred to the Appendix.

\section{Assessing concordance}\label{sec:methods}

In order to quantify the concordance between different cosmological parameter constraints we employ the Surprise statistic \cite{Seehars:2014}. The Surprise $S$ is based on the Kullback-Leibler (KL) divergence or relative entropy \cite{Kullback:1951} $D$ of two probability distribution functions (pdf) $p_{1}(\boldsymbol{\theta})$, $p_{2}(\boldsymbol{\theta})$, which is given by
\begin{equation}
D(p_{2}||p_{1}) = \int\mathrm{d}\boldsymbol{\theta} \: p_{2}(\boldsymbol{\theta}) \log \frac{p_{2}(\boldsymbol{\theta})}{p_{1}(\boldsymbol{\theta})}.
\label{eq:kldiv}
\end{equation}
The KL-divergence gives a measure of information gain from a prior distribution $p_{1}(\boldsymbol{\theta})$ to a posterior distribution $p_{2}(\boldsymbol{\theta})$ in a Bayesian framework \cite{Cover:2006}.
The Surprise is then defined as the difference between the measured relative entropy $D$ and its expectation value $\langle D \rangle$, i.e. $S = D -\langle D \rangle$. The quantity $\langle D \rangle$ represents the expectation for the posterior based on the prior and the likelihood for the new experiment; for a more detailed description, see Ref.~\cite{Seehars:2014}. By construction, the Surprise is expected to scatter around zero. A large positive Surprise suggests that the posterior is more different from the prior than expected a priori while a negative Surprise suggests an agreement that is better than expected. When the base of the logarithm in Eq.~\ref{eq:kldiv} is 2, the relative entropy and the Surprise are measured in bits.

In cosmology, there are two main situations in which to update prior knowledge. The first case consists of an update of the parameter constraints using uncorrelated or weakly correlated data. In this case we can obtain the updated distribution by simply multiplying the prior distribution with the likelihood of the new data. Following Ref.~\cite{Seehars:2014} we denote this case as `add'. The second case consists in updating parameter constraints using correlated data. Such data sets need to be combined in a joint analysis, which takes the cross-correlations of the data into account. However, if two data sets are strongly correlated and the posterior is superior to the prior, it can be more efficient to replace the prior constraints with the posterior constraints. As in Ref.~\cite{Seehars:2014} we refer to this second case as `replace'. In these two cases, Ref.~\cite{Seehars:2014} showed that the Surprise can be computed analytically from the moments of samples of the pdfs when assuming all distributions to be Gaussian, and the model to be linear in the parameters. As shown in Ref.~\cite{Seehars:2014}, these are reasonable approximations for CMB data sets. In these limits, the Surprise for an update from a prior $p_{1}(\boldsymbol{\theta}) = \mathcal{N}(\boldsymbol{\theta}, \boldsymbol{\theta}_{1}, \Sigma_{1})$ to a posterior $p_{2}(\boldsymbol{\theta}) = \mathcal{N}(\boldsymbol{\theta}, \boldsymbol{\theta}_{2}, \Sigma_{2})$ is given by \cite{Seehars:2014}
\begin{equation}
\begin{aligned}
S &= D - \langle D \rangle = \frac{1}{2}(\boldsymbol{\theta}_{2}-\boldsymbol{\theta}_{1})^{\mathrm{T}}\Sigma^{-1}_{1}(\boldsymbol{\theta}_{2}-\boldsymbol{\theta}_{1}) + \frac{1}{2}[\mathrm{tr}(\Sigma^{-1}_{1}\Sigma_{2}) - d] \; \; \; \; \; \mathrm{add}, \\
S &= D - \langle D \rangle = \frac{1}{2}(\boldsymbol{\theta}_{2}-\boldsymbol{\theta}_{1})^{\mathrm{T}}\Sigma^{-1}_{1}(\boldsymbol{\theta}_{2}-\boldsymbol{\theta}_{1}) - \frac{1}{2}[\mathrm{tr}(\Sigma^{-1}_{1}\Sigma_{2}) + d] \; \; \; \; \;  \mathrm{replace},
\label{eq:entropy}
\end{aligned}
\end{equation}
where $ \mathcal{N}(\boldsymbol{\theta}, \boldsymbol{\theta}_{i}, \Sigma_{i})$ denotes a $d$-dimensional Gaussian distribution with mean $\boldsymbol{\theta}_{i}$ and covariance matrix $\Sigma_{i}$.
The relative entropy $D$ is expected to fluctuate around $\langle D \rangle$ with a variance given by \cite{Seehars:2014}
\begin{equation}
\sigma^{2}(D) = \frac{1}{2}\mathrm{tr}((\Sigma^{-1}_{1}\Sigma_{2} \pm \mathbbm{1} )^{2}), 
\end{equation}
where the $+$ holds when replacing correlated data, while the $-$ holds when adding complementary data. The relative entropy $D$ follows a generalised chi-squared ($\chi^{2}$) distribution, which allows us to compute the significance of the obtained Surprise as shown in Ref.~\cite{Seehars:2014}.

In order for Eq.~\ref{eq:entropy} to be a reasonable approximation to the true Surprise, the distributions considered need to be well-approximated by multivariate Gaussians. Assessing Gaussianity can be complicated in high-dimensional parameter space and we therefore choose to Gaussianise all samples before computing the Surprise with Eq.~\ref{eq:entropy}. A common Gaussianisation method is the so-called Box-Cox transformation \cite{Box:1964}. In this work, we consider the one-parameter Box-Cox transformation, which for a sample of a $d$-dimensional parameter set $\boldsymbol{\theta}=(\theta_{ik}), \; i \in [1, n_{\mathrm{samp}}], \; k \in [1, d]$ is defined as \cite{Box:1964}:
\begin{equation}
\theta_{ik}^{'(\lambda_{k})} = \begin{cases} \lambda_{k}^{-1}(\theta_{ik}^{\lambda_{k}}-1) & \mathrm{if} \; \lambda_{k} \neq 0, \\ 
								     \log(\theta_{ik}) & \mathrm{if} \; \lambda_{k}=0.
						\end{cases} 
\label{eq:boxcox}
\end{equation}
This transformation is defined only for $\theta_{ik} > 0$. The parameters $\lambda_{k}$ denote the Box-Cox parameters that render the 1D distribution of the transformed $\theta_{ik}^{'(\lambda_{k})}$ approximately Gaussian. To perform all Box-Cox transformations, we use the $\tt{python}$ library $\tt{scipy}$\footnote{Specifically, we use the $\tt{scipy}$ method $\tt{scipy.stats.boxcox}$.}. In this implementation, the optimal parameters $\lambda_{k}$ are determined by maximising the likelihood for the transformed sample to be Gaussian. For a more detailed discussion of Box-Cox transformations the reader is referred to Refs.~\cite{Box:1964, Velilla:1993, Joachimi:2011, Schuhmann:2016}.

Since even the optimal Box-Cox transformation does not always result in sufficiently Gaussian distributions we follow a similar approach to Refs.~\cite{Schuhmann:2016, Grandis:2016aa} and iteratively Gaussianise the chains. We compute the Surprise for an update from the prior $p_{1}(\boldsymbol{\theta})$ to the posterior $p_{2}(\boldsymbol{\theta})$ from the respective pdf samples $\boldsymbol{\theta}^{1}=(\theta^{1}_{ij}), \; i \in [1, n_{\mathrm{samp, 1}}], \; j \in [1, d]$ and $\boldsymbol{\theta}^{2}=(\theta^{2}_{ij}), \; i \in [1, n_{\mathrm{samp, 2}}], \; j \in [1, d]$ with the procedure outlined in Algorithm \ref{alg:box-cox}.

\begin{algorithm}
\caption{Algorithm used to compute the relative entropy.}\label{alg:box-cox}
\begin{algorithmic} 
\FOR{$i \leq N_{\mathrm{it}}$}
\IF{$i > 1$}
\STATE Determine the linear transformation $\psi$ that standardises the prior sample $\boldsymbol{\theta}^{1}$.
\STATE $\boldsymbol{\theta}^{1'} \leftarrow \psi(\boldsymbol{\theta}^{1})$
\STATE Apply $\psi$ to the posterior sample $\boldsymbol{\theta}^{2}$.
\STATE $\boldsymbol{\theta}^{2'} \leftarrow \psi(\boldsymbol{\theta}^{2})$
\ENDIF
\STATE Transform to positive parameter values
\FOR{$k \leq d$}
\STATE $a_{k} = \min_{ll'}(\boldsymbol{\theta}^{1'}_{lk}, \boldsymbol{\theta}^{2'}_{l'k})$
\IF{$a_{k} < 0$}
\STATE $\boldsymbol{\theta}^{1'}_{lk} \leftarrow \boldsymbol{\theta}^{1'}_{lk} - a_{k}, \; l \in [1, n_{\mathrm{samp, 1}}]$
\STATE $\boldsymbol{\theta}^{2'}_{lk} \leftarrow \boldsymbol{\theta}^{2'}_{lk} - a_{k}, \; l \in [1, n_{\mathrm{samp, 2}}]$
\ENDIF
\ENDFOR
\STATE Find the optimal one-parameter Box-Cox transformation $\phi$ for $\boldsymbol{\theta}^{1'}$ (Eq.~\ref{eq:boxcox}).
\STATE $\boldsymbol{\theta}^{1''} \leftarrow \phi(\boldsymbol{\theta}^{1'})$
\STATE Apply the transformation $\phi$ to $\boldsymbol{\theta}^{2'}$
\STATE $\boldsymbol{\theta}^{2''} \leftarrow \phi(\boldsymbol{\theta}^{2'})$
\STATE Compute $D, \langle D \rangle, S, \sigma(D)$ for $\boldsymbol{\theta}^{1''}, \boldsymbol{\theta}^{2''}$.
\ENDFOR
\end{algorithmic}
\end{algorithm}

In the first step, we linearly transform $\boldsymbol{\theta}^{1}$ into the eigenbasis of its correlation matrix, which yields the linear transformation $\psi$. The transformed samples $\boldsymbol{\theta}^{1'}$ are therefore approximately uncorrelated and have zero mean and unit standard deviation (see e.g. Ref.~\cite{Seehars:2016}). In the second step we apply the same transformation $\psi$ to the samples $\boldsymbol{\theta}^{2}$; the results of this transformation are denoted as $\boldsymbol{\theta}^{2'}$\footnote{We do not perform the standardisation step in the first iteration, since it can result in significantly non-Gaussian constraints. This can cause difficulties in the convergence of our algorithm or it can even cause it to fail. Nevertheless, in order to obtain numerically stable results for the update CMB set $\rightarrow$ P15 in the $\Lambda$CDM parametrisation with $\sigma_{8}$ we do perform the standardisation step in the first iteration, because the algorithm fails otherwise. We will come back to this case in Sec.~\ref{sec:results}.}.  As noted above, the Box-Cox transformation is only defined for positive values. In step 3, we therefore shift the new parameters such that each value is positive. We repeat this procedure for $N_{\mathrm{it}}$ iterations while monitoring the values of $D, \langle D \rangle, S$ and $\sigma(D)$. We terminate the iteration as soon as the entropy values reach a steady state. In order to compute the entropy values we use the publicly available $\texttt{Surprise}$ package\footnote{$\texttt{https://github.com/seeh/surprise}$} described in Ref.~\cite{Seehars:2016}. We monitor the Gaussianity of the chains in each iteration by testing if the Mahalanobis distance \cite{Mahalanobis:1936} of the samples is distributed as a $\chi^{2}$ variable \cite{Mardia:1979, Gnanadesikan:1972, Wilks:1963}, which is a good approximation for the sample sizes used in this work \cite{Joenssen:2014}. In the Appendix we show an example of the performed Gaussianity tests.

Using the above prescription, we compute the Surprise between the different data sets for a fiducial parametrisation of the $\Lambda$CDM cosmological model given by $\{h,\allowbreak\, \Omega_{\mathrm{c}}h^{2},\allowbreak \,\Omega_{\mathrm{b}}h^{2},\allowbreak \, n_{\mathrm{s}},\allowbreak \,A_{\mathrm{s}},\allowbreak \,\tau_{\mathrm{reion}}\}$. In this notation, $h$ is the dimensionless Hubble parameter, $\Omega_{\mathrm{c}}$ is the fractional cold dark matter density today, $\Omega_{\mathrm{b}}$ is the fractional baryon density today, $n_{\mathrm{s}}$ denotes the scalar spectral index, $A_{\mathrm{s}}$ is the primordial power spectrum amplitude at a pivot scale of $k_{0} = 0.05$ Mpc$^{-1}$ and $\tau_{\mathrm{reion}}$ denotes the optical depth to reionisation. In order to assess the stability of our results, we also repeat the above computations for the following two reparametrisations of $\Lambda$CDM: $\{h,\, \Omega_{\mathrm{m}}, \,\Omega_{\mathrm{b}}, \, n_{\mathrm{s}}, \,\sigma_{8}, \,\tau_{\mathrm{reion}}\}$, $\{h,\, \Omega_{\mathrm{m}}, \,\Omega_{\mathrm{b}}, \, n_{\mathrm{s}}, \,A_{\mathrm{s}}, \,\tau_{\mathrm{reion}}\}$,  where $\Omega_{\mathrm{m}}$ denotes the fractional matter density today\footnote{The parameters $\Omega_{\mathrm{m}}$ and $\Omega_{\mathrm{c}}$ are related by $\Omega_{\mathrm{m}}=\Omega_{\mathrm{c}}+\Omega_{\mathrm{b}}+\Omega_{\mathrm{ncdm}}$, where $\Omega_{\mathrm{ncdm}}$ is the fractional non-cold dark matter density today.} and $\sigma_{8}$ is the r.m.s. of linear matter fluctuations in spheres of comoving radius $8 \,h^{-1}$ Mpc. This allows us to test the independence of the Surprise statistic on the parametrisation of the cosmological model \cite{Kullback:1951, Seehars:2014} as well as to test if the assumptions of Gaussianity of the pdf samples and linearity of the model parameters are appropriate. An additional way to test the Gaussianity of the CMB constraints is to make use of the so-called CMB Gaussian parameters derived in Ref.~\cite{Chu:2003}: these parameters exhibit an approximately Gaussian likelihood and are obtained through a nonlinear transformation of the cosmological parameters. Explicitly, they are given by $\{\Omega_{\mathrm{c}}h^{2},\allowbreak \,\Omega_{\mathrm{b}}h^{2},\allowbreak \, \theta_{*}, \allowbreak A_{\mathrm{s}} e^{-2\tau_{\mathrm{reion}}}, \allowbreak z_{\mathrm{reion}}, \allowbreak \sfrac{1}{\sqrt{\Omega_{\mathrm{b}}h^{2}}} 2^{n_{\mathrm{s}}-1}\}$, where $z_{\mathrm{reion}}$ is the reionisation redshift and $\theta_{*}$ denotes the angular size of the sound horizon at the redshift $z_{*}$ for which the optical depth in the absence of reionisation equals unity. Since these parameters are approximately Gaussian distributed for CMB data \cite{Chu:2003}, the relative entropies in this parameter set should be well approximated by Eq.~\ref{eq:entropy}. When comparing CMB data sets we therefore additionally compute the relative entropies for the CMB Gaussian parameters\footnote{We note that these parameters were not available to us for P15$^{*}$ and we therefore did not perform this additional test for CMB comparisons involving P15$^{*}$.}.

\section{Data}\label{sec:data}

In our analysis, we consider the constraints derived from the integrated analysis of Ref.~\cite{Nicola:2017}, the constraints derived by the Planck Collaboration \cite{Planck-Collaboration:2016ae} (TT+lowP) in their second data release as well as those derived from the combination of WMAP9 data \cite{Hinshaw:2013} with high-$\ell$ data from ACT \cite{Fowler:2010, Das:2011} and SPT \cite{Keisler:2011, Reichardt:2012}. We further include the recent update to the WMAP9+SPT+ACT constraints from Ref.~\cite{Calabrese:2017}. For all these data sets except the IA, we use the publicly available parameter chains from the \textsc{Planck Legacy Archive}\footnote{\texttt{http://pla.esac.esa.int/pla/\#cosmology}} and \textsc{Lambda}\footnote{\texttt{https://lambda.gsfc.nasa.gov/product/}}.

We further complement these data sets by computing cosmological parameter constraints in the framework of a $\Lambda$CDM cosmological model from the publicly available Planck CMB temperature likelihood for $\ell \in [2, 2508]$. We sample the likelihood in a Monte Carlo Markov Chain (MCMC) with $\texttt{CosmoHammer}$ \cite{Akeret:2013} and vary the six cosmological parameters $\{h,\allowbreak \, \Omega_{\mathrm{c}},\allowbreak \,\Omega_{\mathrm{b}},\allowbreak \, n_{\mathrm{s}},\allowbreak \,A_{\mathrm{s}},\allowbreak \,\tau_{\mathrm{reion}}\}$. In addition to the cosmological parameters, we also vary the 15 nuisance parameters employed by the Planck Collaboration \cite{Planck-Collaboration:2016ae} along with the respective Gaussian priors. Since the optical depth to reionisation $\tau_{\mathrm{reion}}$ cannot be constrained from only CMB temperature data, we follow Ref.~\cite{Calabrese:2017} and include the recent Planck measurement of $\tau_{\mathrm{reion}}$ \cite{Planck-Collaboration:2016aa} by assuming a Gaussian prior of $\tau_{\mathrm{reion}} = 0.06 \pm 0.01$. For the computation of theoretical CMB power spectra, we use the Boltzmann code $\texttt{class}$\footnote{$\tt{http://class\text{-}code.net}$.} \cite{Blas:2011, Lesgourgues:2011}. Following Ref.~\cite{Planck-Collaboration:2016ae}, our fiducial model contains two massless neutrinos and one massive neutrino with the fixed minimal mass of $\sum m_{\nu} = 0.06$ eV.  

In order to assess the consistency between the constraints derived in Ref.~\cite{Nicola:2017} and those from the Planck Collaboration \cite{Planck-Collaboration:2016ae}, we further compute constraints from the IA combined with Planck CMB temperature data for $\ell \in [802, 2508]$. Since the IA contains Planck CMB temperature data up to a maximal angular multipole $\ell_{\mathrm{max}} = 610$ and cross-correlations between CMB temperature anisotropies and other cosmological probes are mostly confined to large angular scales, we assume these two datasets to be independent and perform importance sampling to derive the joint constraints. In practice, we first compute the value of the Planck likelihood for all the points in the IA chains, fixing the nuisance parameters to their best-fit values derived in Ref.~\cite{Planck-Collaboration:2016ae} (TT+lowP). We then use these likelihood values as weights for the IA chains, which results in chains weighted according to the product of both the IA and the high-$\ell$ Planck likelihood. 

All these constraints are derived assuming flat priors on all cosmological parameters except the optical depth to reionisation $\tau_{\mathrm{reion}}$. These priors are not identical for different data sets but they are all significantly broader than the obtained posterior constraints and they can therefore be approximated as equal over the support of the posterior distributions. A summary of the data sets used in this work can be found in Tab.~\ref{tab:datasets}. 

\begin{table*}
\caption{Overview of the data sets considered in this work.} \label{tab:datasets}
\begin{center}
\begin{tabular}{>{\centering}m{2.6cm}>{\centering}m{8cm}>{\centering}m{2.5cm}@{}m{0pt}@{}} \hline\hline 
Acronym & Description & Reference & \\ \hline \Tstrut     
IA & Integrated analysis with WMAP9 $\tau_{\mathrm{reion}}$-prior, $\tau_{\mathrm{reion}} = 0.089 \pm 0.02$ ($\sum m_{\nu} = 0.0$ eV) & \cite{Nicola:2016, Nicola:2017} & \\ \\
IA* & Integrated analysis with Hubble parameter measurement from R16, \\$H_{0} = 73.24 \pm1.74$ km s$^{-1}$ Mpc$^{-1}$ & \cite{Riess:2016, Nicola:2016, Nicola:2017} & \\ \\
P15 & Constraints from Planck Collaboration 2015, TT+lowP ($\sum m_{\nu} = 0.06$ eV) & \cite{Planck-Collaboration:2016ae} & \\ \\
CMB set & Constraints from WMAP9+SPT+ACT ($\sum m_{\nu} = 0.0$ eV) & \cite{Hinshaw:2013} & \\ \\
CMB set$^{*}$ & Updated constraints from WMAP9+SPT+ACT \\ with Planck HFI $\tau_{\mathrm{reion}}$-prior, \\$\tau_{\mathrm{reion}} = 0.06 \pm 0.01$ ($\sum m_{\nu} = 0.06$ eV). & \cite{Calabrese:2017, Planck-Collaboration:2016aa} & \\ \\
P15$^{*}$ & Constraints from Planck TT likelihood \\ for $\ell \in [2, 2508]$ and Planck HFI $\tau_{\mathrm{reion}}$-prior, \\$\tau_{\mathrm{reion}} = 0.06 \pm 0.01$ ($\sum m_{\nu} = 0.06$ eV). & \cite{Planck-Collaboration:2016ae, Planck-Collaboration:2016aa} & \\ \\
IA+P15$^{**}$ hi-$\ell$, massless $\nu$ & Constraints from combination of IA with Planck TT likelihood for $\ell \in [802, 2508]$ ($\sum m_{\nu} = 0.0$ eV). & This work, \cite{Nicola:2016, Nicola:2017, Planck-Collaboration:2016ae} & \\ \\
IA+P15$^{**}$ hi-$\ell$, massive $\nu$ & Constraints from combination of IA with Planck TT likelihood for $\ell \in [802, 2508]$ ($\sum m_{\nu} = 0.06$ eV). & This work, \cite{Nicola:2016, Nicola:2017, Planck-Collaboration:2016ae} & \\ \hline \hline
\end{tabular}
\end{center}
\end{table*} 

\section{Results}\label{sec:results}

\begin{figure*}
\begin{center}
\includegraphics[scale=0.45]{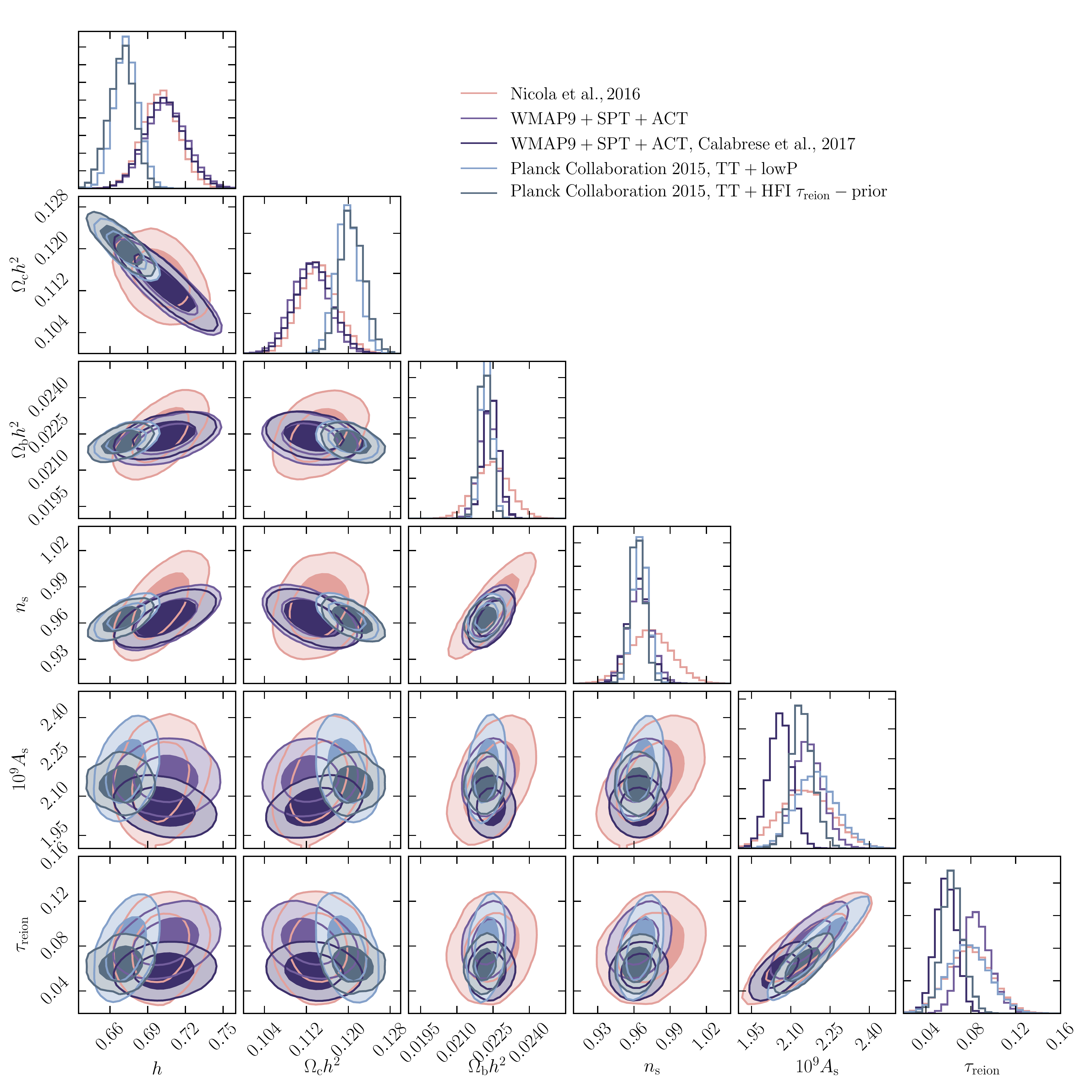}
\caption{Comparison of the constraints obtained from the integrated analysis to the constraints obtained by the Planck Collaboration (TT+lowP), the constraints obtained from the Planck TT likelihood with the HFI $\tau_{\mathrm{reion}}$-prior, the constraints obtained by WMAP9+SPT+ACT and the updated constraints from WMAP9+SPT+ACT \cite{Calabrese:2017} in the $\Lambda$CDM parametrisation $\{h,\, \Omega_{\mathrm{c}}h^{2}, \,\Omega_{\mathrm{b}}h^{2}, \, n_{\mathrm{s}}, \,A_{\mathrm{s}}, \,\tau_{\mathrm{reion}}\}$. All constraints are marginalised over the respective nuisance parameters. In each case the inner (outer) contour shows the $68 \%$ c.l. ($95 \%$ c.l.).}
\label{fig:constraints-comp-As-omegach2-omegabh2}
\end{center}
\end{figure*}

\begin{figure*}
\begin{center}
\includegraphics[scale=0.45]{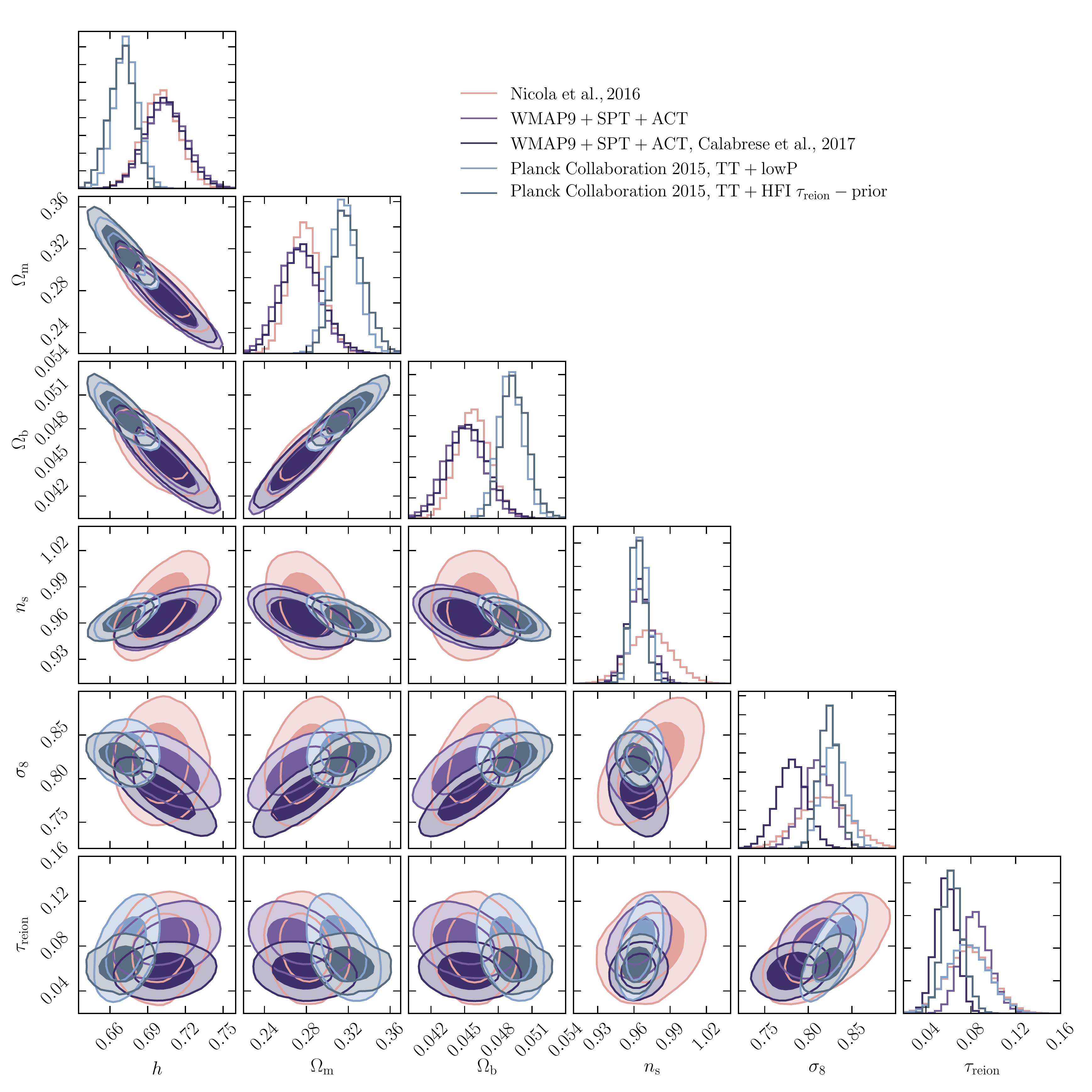}
\caption{Same as Fig.~\ref{fig:constraints-comp-As-omegach2-omegabh2} but showing the constraints on the parameters $\{h,\, \Omega_{\mathrm{m}}, \,\Omega_{\mathrm{b}}, \, n_{\mathrm{s}}, \,\sigma_{8}, \,\tau_{\mathrm{reion}}\}$ instead of $\{h,\, \Omega_{\mathrm{c}}h^{2}, \,\Omega_{\mathrm{b}}h^{2}, \, n_{\mathrm{s}}, \,A_{\mathrm{s}}, \,\tau_{\mathrm{reion}}\}$.}
\label{fig:constraints-comp-sigma8-omegam}
\end{center}
\end{figure*}

Using the procedure outlined in Sec.~\ref{sec:methods}, we compute the estimates of the relative entropy for the different data sets considered. We compute all quantities from moments of the Monte Carlo Markov chain samples, choosing a fiducial $\Lambda$CDM parametrisation of $\{h,\allowbreak\, \Omega_{\mathrm{c}}h^{2},\allowbreak \,\Omega_{\mathrm{b}}h^{2},\allowbreak \, n_{\mathrm{s}},\allowbreak \,A_{\mathrm{s}},\allowbreak \,\tau_{\mathrm{reion}}\}$. Where possible, we make use of the publicly available parameter chains. The numerical values for the estimated entropies are given in Tab.~\ref{tab:relent}, where we show the entropies computed in our fiducial $\Lambda$CDM parametrisation together with the results for $\{h,\allowbreak\, \Omega_{\mathrm{m}},\allowbreak \,\Omega_{\mathrm{b}},\allowbreak \, n_{\mathrm{s}},\allowbreak \,\sigma_{8},\allowbreak \,\tau_{\mathrm{reion}}\}$ in parentheses. For the first two comparisons in the table we further show the results obtained using the CMB Gaussian parameters $\{\Omega_{\mathrm{c}}h^{2},\allowbreak \,\Omega_{\mathrm{b}}h^{2},\allowbreak \, \theta_{*}, \allowbreak A_{\mathrm{s}} e^{-2\tau_{\mathrm{reion}}}, \allowbreak z_{\mathrm{reion}}, \allowbreak \sfrac{1}{\sqrt{\Omega_{\mathrm{b}}h^{2}}} 2^{n_{\mathrm{s}}-1}\}$ \cite{Chu:2003} without performing any Gaussianisation. The constraints for a subset of the data sets considered are shown in Figures \ref{fig:constraints-comp-As-omegach2-omegabh2} and \ref{fig:constraints-comp-sigma8-omegam} for the former two parametrisations of the $\Lambda$CDM cosmological model.

We first consider the relative entropy between WMAP9+SPT+ACT (denoted CMB set) \cite{Hinshaw:2013} and Planck 2015 (TT+lowP) (denoted P15) \cite{Planck-Collaboration:2016ae}. These two data sets are strongly correlated with one another, especially on large angular scales. We therefore replace the WMAP9+SPT+ACT constraints with those from Planck 2015 (TT+lowP) when computing the relative entropy. As can be seen from Tab.~\ref{tab:relent} we find a negative Surprise ($S = -2.3$, $\sigma(D)=5.5$), suggesting a slightly better agreement between the two data sets than expected a priori. This was also seen in Ref.~\cite{Seehars:2016} for a replacement of WMAP9 with Planck 2015 temperature and polarisation data. When computing the relative entropy in the $\Lambda$CDM parametrisation with $\sigma_{8}$ on the other hand, we find significantly different results ($S = 5.1$, $\sigma(D) = 4.5$). This appears surprising at first, since the relative entropy is invariant under transformations of the model parameters. Furthermore, the computation of the Surprise obtained with $\sigma_{8}$ becomes numerically unstable. This behaviour is due to the differences in the fiducial neutrino model assumed by WMAP9+SPT+ACT and Planck 2015: the WMAP9+SPT+ACT constraints are derived assuming massless neutrinos, whereas the fiducial model used by the Planck Collaboration includes two massless and one massive neutrino with $\sum m_{\nu} = 0.06$ eV. Being a parameter of the early Universe, the constraints on $A_{\mathrm{s}}$ are not significantly affected by a different neutrino model. The constraints on $\sigma_{8}$ on the other hand are sensitive to the mapping between early and late Universe and are therefore significantly affected by differences in the neutrino model. When we compare the constraints from WMAP9+SPT+ACT and Planck 2015 in $\sigma_{8}$, we therefore compare two different fiducial models with each other and this is the reason for the discussed instabilities in the relative entropy (see also Sec.~\ref{sec:methods}). In other words, the relative entropy is not invariant under inconsistent transformations of the considered prior and posterior and is therefore sensitive to inconsistencies in the models that are used in different parts of the analysis. As discussed in Sec.~\ref{sec:methods}, we additionally compute the relative entropies using the CMB Gaussian parameters \cite{Chu:2003} in order to test the Gaussianisation method. The results obtained are similar to those obtained in our fiducial $\Lambda$CDM parameterisation, thus suggesting that the procedure described in Sec.~\ref{sec:methods} yields chains sufficiently Gaussian for our purpose.

Ref.~\cite{Calabrese:2017} recently computed updated WMAP9+SPT+ACT constraints (denoted CMB set$^{*}$) assuming a single family of massive neutrinos with $\sum m_{\nu} = 0.06$ eV. We compare these constraints to those from Planck 2015 (TT+lowP) by computing the relative entropy when replacing the revised WMAP9+SPT+ACT constraints with those from Planck 2015.  As can be seen from Tab.~\ref{tab:relent}, we find these two data sets to be in good agreement ($S = 1.0$, $\sigma(D) = 6.4$), consistent with the results presented in Ref.~\cite{Calabrese:2017}. Furthermore, the results are insensitive to the parametrisation of the $\Lambda$CDM cosmological model. This confirms that the discrepancies discussed before are due to the different neutrino models assumed and additionally confirms the Gaussianisation procedure chosen in this work. We finally compare the updated WMAP9+SPT+ACT constraints from Ref.~\cite{Calabrese:2017} to the constraints obtained from the Planck TT likelihood including the recent Planck High Frequency Instrument (HFI) $\tau_{\mathrm{reion}}$ measurement (P15$^{*}$). When computing the relative entropy for a replacement of the constraints from Ref.~\cite{Calabrese:2017} with the Planck TT likelihood constraints, we find a minor improvement in the agreement between the two data sets ($S = -0.1$, $\sigma(D) = 4.7$). These results are independent of the $\Lambda$CDM parametrisation, as expected.

Finally, we also assess the consistency between the constraints derived in the IA with those derived by the Planck Collaboration \cite{Planck-Collaboration:2016ae}. To this end, we estimate the relative entropy between the IA constraints \cite{Nicola:2016, Nicola:2017} and the constraints obtained after adding Planck high-$\ell$ data to the IA through importance sampling (denoted IA+P15$^{**}$ hi-$\ell$). In order to investigate the impact of the neutrino model on the consistency between these two data sets, we once weight the IA chains by the Planck likelihood values computed assuming massless neutrinos (as we did in the IA), and once with the Planck likelihood values computed assuming the fiducial Planck neutrino model. When we consistently assume massless neutrinos we find a good agreement between the constraints from these two data sets ($S = 0.1$, $\sigma(D) = 1.7$). Furthermore, we find the results to be independent of the parametrisation of the cosmological model, as can be seen from Tab.~\ref{tab:relent}. In the case in which we assume different neutrino models, we still find consistency between the two data sets in our fiducial parametrisation ($S = 0.7$, $\sigma(D) = 1.7$), but we cannot obtain stable results for the $\sigma_{8}$ parametrisation of $\Lambda$CDM. Similar to before, we thus find that the Surprise depends on the parametrisation of the cosmological model only for inconsistent comparisons. We stress that constraints derived using different fiducial models should not be compared and the cases considered here should therefore serve as a warning illustrating the effects of inconsistent comparisons.

In the IA we have used the value of the Hubble parameter derived by Ref.~\cite{Efstathiou:2014} (denoted E14), which is a reanalysis of the measurement by Ref.~\cite{Riess:2011}. This measurement has been updated with a more precise value in Ref.~\cite{Riess:2016} (denoted R16), which has been found to be in tension with the results from Ref.~\cite{Planck-Collaboration:2016ae} (see e.g. Refs.~\cite{Grandis:2016aa, Heavens:2017}). In order to investigate the impact of this choice on the consistency between the constraints derived in the IA and those derived by the Planck Collaboration \cite{Planck-Collaboration:2016ae}, we additionally compute the relative entropy between the IA combined with R16 (denoted IA*) and the constraints obtained when combining the IA* with Planck high-$\ell$ data. As can be seen from Tab.~\ref{tab:relent} we find the two data sets to be in agreement with each other, albeit with a slightly lower $p$-value.

\begin{table*}
\caption{Values of the relative entropy $D$, expected relative entropy $\langle D \rangle$, standard deviation of the relative entropy $\sigma(D)$ and Surprise $S$ for the parameter updates considered in this work. All the values are given in units of bits. The $p$-value denotes the probability of observing a value of the Surprise that is greater or equal (less or equal) than $S$ if $S$ is greater (smaller) than zero when assuming consistency between the two data sets. The fiducial entropy values are computed for the $\Lambda$CDM parametrisation $\{h,\allowbreak\, \Omega_{\mathrm{c}}h^{2},\allowbreak \,\Omega_{\mathrm{b}}h^{2},\allowbreak \, n_{\mathrm{s}},\allowbreak \,A_{\mathrm{s}},\allowbreak \,\tau_{\mathrm{reion}}\}$, while the values in parentheses show the results for $\{h,\allowbreak\, \Omega_{\mathrm{m}},\allowbreak \,\Omega_{\mathrm{b}},\allowbreak \, n_{\mathrm{s}},\allowbreak \,\sigma_{8},\allowbreak \,\tau_{\mathrm{reion}}\}$ (top) and $\{\Omega_{\mathrm{c}}h^{2},\allowbreak \,\Omega_{\mathrm{b}}h^{2},\allowbreak \, \theta_{*}, \allowbreak A_{\mathrm{s}} e^{-2\tau_{\mathrm{reion}}}, \allowbreak z_{\mathrm{reion}}, \allowbreak \sfrac{1}{\sqrt{\Omega_{\mathrm{b}}h^{2}}} 2^{n_{\mathrm{s}}-1}\}$ (bottom) respectively.} \label{tab:relent}
\begin{center}
\begin{tabular}{>{\raggedright}m{1.7cm}c>{\centering}m{2.6cm}>{\centering}m{1.4cm}ccccc}
\hline\hline 
\multicolumn{3}{l}{Data combination} & Updating \\ scheme & $D$ & $\langle D \rangle$ & $S$ & $\sigma(D)$ & $p$-value \\ \hline      
\multirow{3}{*}{CMB set}& \multirow{3}{*}{$\rightarrow$} & \multirow{3}{*}{P15} & \multirow{3}{*}{replace} & 9.1 & 11.5 & -2.3 & 5.5 & 0.4 \\
 & & & & (17.5) & (12.4) &  (5.1) & (4.5) & (0.1) \\
  & & & & (8.3) & (11.6) & (-3.3) & (5.5) & (0.3) \\\hdashline
\multirow{3}{*}{CMB set$^{*}$} & \multirow{3}{*}{$\rightarrow$} & \multirow{3}{*}{P15} & \multirow{3}{*}{replace} & 13.7 & 12.6 & 1.1 & 6.5 & 0.3 \\ 
 & & & & (13.1) & (13.4) & (-0.4)& (7.1) & (0.6) \\
  & & & & (12.6) & (12.1) & (0.5)& (6.1) & (0.4) \\\hdashline
\multirow{2}{*}{CMB set$^{*}$ }& \multirow{2}{*}{$\rightarrow$} & \multirow{2}{2.5cm}{ \centering P15$^{*}$} & \multirow{2}{*}{replace} & 9.6 & 9.7 & -0.1 & 4.7 & 0.6 \\
 & & & & (10.1) & (10.4) & (-0.3)&(5.2) & (0.6) \\ \hline 
\multirow{2}{*}{IA} & \multirow{2}{*}{$\rightarrow$} & \multirow{2}{2.6cm}{\centering IA+P15$^{**}$ hi-$\ell$, massless $\nu$} & \multirow{2}{*}{add} & 8.3 & 8.2 & 0.1 & 1.7 & 0.4 \\
 & & & & (8.3) & (8.1) & (0.2)&(1.7) & (0.4) \\\hdashline
 \multirow{2}{*}{IA*} & \multirow{2}{*}{$\rightarrow$} & \multirow{2}{2.8cm}{\centering IA*+P15$^{**}$ hi-$\ell$, massless $\nu$} & \multirow{2}{*}{add} & 10.7 & 9.3 & 1.5 & 1.8 & 0.2 \\
 & & & & (10.6) & (9.5) & (1.2)&(1.9) & (0.2) \\\hdashline
\multirow{2}{*}{IA} & \multirow{2}{*}{$\rightarrow$} & \multirow{2}{2.6cm}{\centering IA+P15$^{**}$ hi-$\ell$, massive $\nu$} & \multirow{2}{*}{add} & 8.9 & 8.2 & 0.7 & 1.7 & 0.3 \\
 & & & & (-) & (-) & (-) & (-) & (-) \\
 \hline \hline
\end{tabular}
\end{center}
\end{table*}

\section{Conclusions}\label{sec:conclusions}

In this work, we have used the Surprise statistic introduced in Ref.~\cite{Seehars:2014} to quantify the concordance between different data sets in the framework of a $\Lambda$CDM cosmological model. We have compared the constraints obtained from the integrated analysis presented in Refs.~\cite{Nicola:2016, Nicola:2017} to those obtained by the Planck Collaboration \cite{Planck-Collaboration:2016ae}. We have further quantified the agreement between the constraints obtained from CMB data by the Planck Collaboration \cite{Planck-Collaboration:2016ae}, WMAP9+SPT+ACT \cite{Hinshaw:2013} and the recently updated WMAP9+SPT+ACT constraints \cite{Calabrese:2017}.

We find good agreement between the different data sets considered, i.e. we find that all the considered CMB data sets yield constraints consistent at a level of less than 1$\sigma$ and we furthermore find that the constraints obtained from the IA are in good agreement with those obtained by the Planck Collaboration. The only apparent discrepancies detected are due to differences in the fiducial neutrino model employed to compute the Planck 2015 and the WMAP9+SPT+ACT constraints and they only appear in specific parametrisations of the cosmological model. Therefore, the relative entropy and the Surprise present a reliable measure of concordance between data sets, as they are sensitive to subtle effects such as differences in neutrino models. These results show that the data sets considered in this work are all well fit by $\Lambda$CDM and do not provide evidence for tensions. 

\acknowledgments

We would like to thank Sebastian Seehars and Sebastian Grandis for useful discussions. We also thank our referee, Stephen Feeney, for the many suggestions that helped us improve the quality and clarity of the manuscript.

\noindent This work was supported in part by SNF grant 200021\_169130.

\noindent This project used public archival data from the Dark Energy Survey (DES). Funding for the DES Projects has been provided by the U.S. Department of Energy, the U.S. National Science Foundation, the Ministry of Science and Education of Spain, the Science and Technology Facilities Council of the United Kingdom, the Higher Education Funding Council for England, the National Center for Supercomputing Applications at the University of Illinois at Urbana-Champaign, the Kavli Institute of Cosmological Physics at the University of Chicago, the Center for Cosmology and Astro-Particle Physics at the Ohio State University, the Mitchell Institute for Fundamental Physics and Astronomy at Texas A$\&$M University, Financiadora de Estudos e Projetos, Funda\c{c}\~{a}o Carlos Chagas Filho de Amparo \`{a} Pesquisa do Estado do Rio de Janeiro, Conselho Nacional de Desenvolvimento Cient\'{i}fico e Tecnol\'{o}gico and the Minist\'{e}rio da Ci\^{e}ncia, Tecnologia e Inova\c{c}\~{a}o, the Deutsche Forschungsgemeinschaft and the Collaborating Institutions in the Dark Energy Survey. The Collaborating Institutions are Argonne National Laboratory, the University of California at Santa Cruz, the University of Cambridge, Centro de Investigaciones En\'{e}rgeticas, Medioambientales y Tecnol\'{o}gicas-Madrid, the University of Chicago, University College London, the DES-Brazil Consortium, the University of Edinburgh, the Eidgen\"{o}ssische Technische Hochschule (ETH) Z\"{u}rich, Fermi National Accelerator Laboratory, the University of Illinois at Urbana-Champaign, the Institut de Ci\`{e}ncies de l'Espai (IEEC/CSIC), the Institut de F\'{i}sica d'Altes Energies, Lawrence Berkeley National Laboratory, the Ludwig-Maximilians Universit\"{a}t M\"{u}nchen and the associated Excellence Cluster Universe, the University of Michigan, the National Optical Astronomy Observatory, the University of Nottingham, the Ohio State University, the University of Pennsylvania, the University of Portsmouth, SLAC National Accelerator Laboratory, Stanford University, the University of Sussex, and Texas A$\&$M University. 

\noindent Funding for the SDSS and SDSS-II has been provided by the Alfred P. Sloan Foundation, the Participating Institutions, the National Science Foundation, the U.S. Department of Energy, the National Aeronautics and Space Administration, the Japanese Monbukagakusho, the Max Planck Society, and the Higher Education Funding Council for England. The SDSS Web Site is http://www.sdss.org/.

\noindent The SDSS is managed by the Astrophysical Research Consortium for the Participating Institutions. The Participating Institutions are the American Museum of Natural History, Astrophysical Institute Potsdam, University of Basel, University of Cambridge, Case Western Reserve University, University of Chicago, Drexel University, Fermilab, the Institute for Advanced Study, the Japan Participation Group, Johns Hopkins University, the Joint Institute for Nuclear Astrophysics, the Kavli Institute for Particle Astrophysics and Cosmology, the Korean Scientist Group, the Chinese Academy of Sciences (LAMOST), Los Alamos National Laboratory, the Max-Planck-Institute for Astronomy (MPIA), the Max-Planck-Institute for Astrophysics (MPA), New Mexico State University, Ohio State University, University of Pittsburgh, University of Portsmouth, Princeton University, the United States Naval Observatory, and the University of Washington.

\noindent Funding for SDSS-III has been provided by the Alfred P. Sloan Foundation, the Participating Institutions, the National Science Foundation, and the U.S. Department of Energy Office of Science. The SDSS-III web site is http://www.sdss3.org/.

\noindent SDSS-III is managed by the Astrophysical Research Consortium for the Participating Institutions of the SDSS-III Collaboration including the University of Arizona, the Brazilian Participation Group, Brookhaven National Laboratory, Carnegie Mellon University, University of Florida, the French Participation Group, the German Participation Group, Harvard University, the Instituto de Astrofisica de Canarias, the Michigan State/Notre Dame/JINA Participation Group, Johns Hopkins University, Lawrence Berkeley National Laboratory, Max Planck Institut for Astrophysics, Max Planck Institute for Extraterrestrial Physics, New Mexico State University, New York University, Ohio State University, Pennsylvania State University, University of Portsmouth, Princeton University, the Spanish Participation Group, University of Tokyo, University of Utah, Vanderbilt University, University of Virginia, University of Washington, and Yale University.

\noindent Based on observations obtained with Planck (http://www.esa.int/Planck), an ESA science mission with instruments and contributions directly funded by ESA Member States, NASA, and Canada.

\noindent The colour palettes employed in this work are taken from $\tt{http://colorpalettes.net}$. The contour plots have been created using $\tt{corner.py}$ \cite{ForemanMackey:2016}. 

\appendix

\section{Gaussianity tests}\label{ap:gaussianity-tests}

\begin{figure*}
\begin{center}
\includegraphics[scale=0.45]{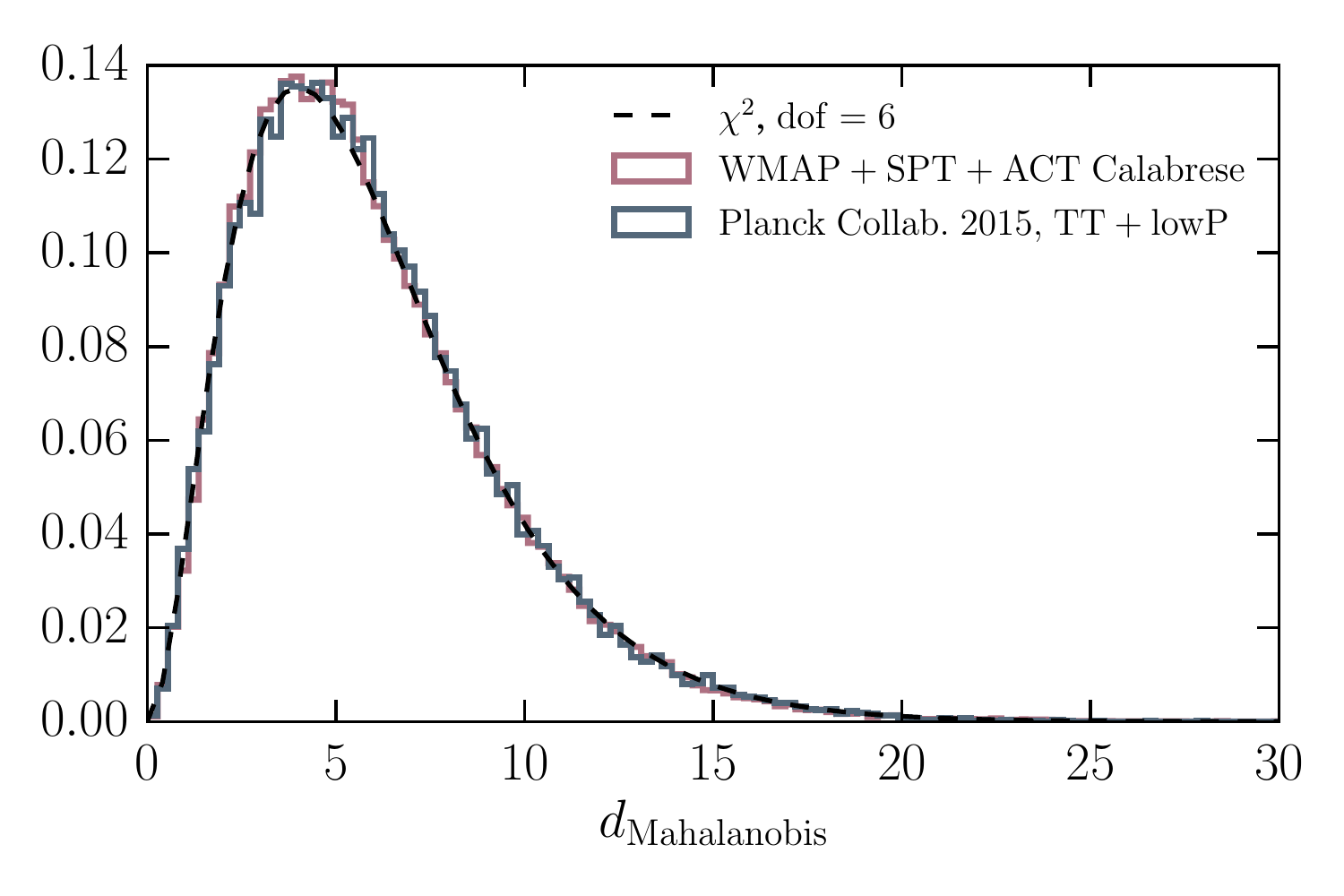}
\caption{Distribution of Mahalanobis distances for Gaussianised CMB set$^{*}$ and P15$^{*}$ chains compared to the theoretically expected $\chi^{2}$ distribution for 6 degrees of freedom.}
\label{fig:mahalanobis-histogram}
\end{center}
\end{figure*}

\begin{figure*}
\begin{center}
\includegraphics[scale=0.45]{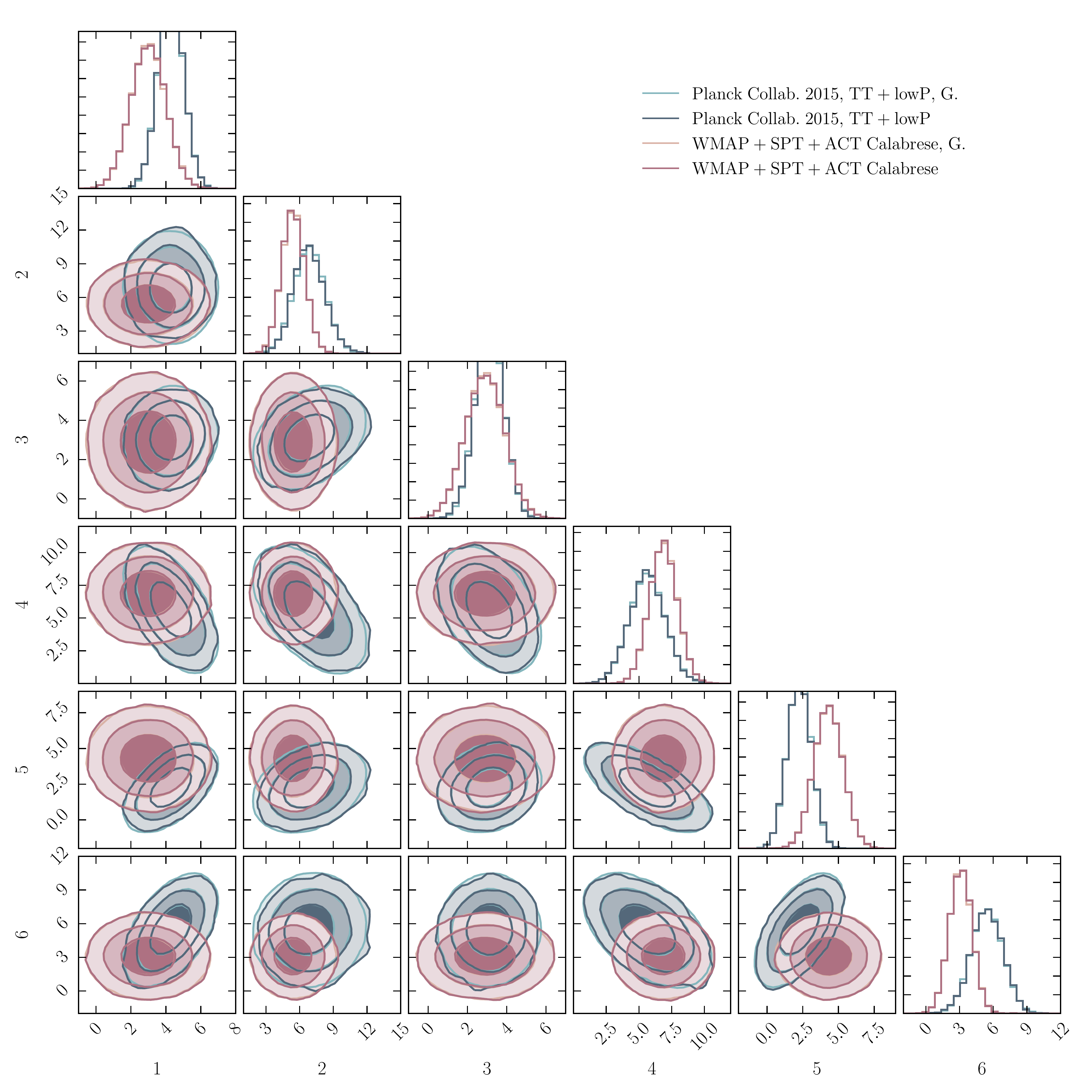}
\caption{Comparison of the Gausssianised contours for CMB set$^{*}$ and P15$^{*}$ to their Gaussian approximations in the transformed parameter space. In each case the first (second, third) contour shows the $68 \%$ c.l. ($95 \%$ c.l., $99 \%$ c.l.).}
\label{fig:contours-vs-gauss}
\end{center}
\end{figure*}

As discussed in Sec.~\ref{sec:methods}, we assess the Gaussianity of the chains during and after the Box-Cox iteration by testing if the Mahalanobis distances \cite{Mahalanobis:1936} of the samples follow a $\chi^{2}$ distribution \cite{Mardia:1979, Gnanadesikan:1972, Wilks:1963}. In this Appendix we consider the Bayesian update CMB set$^{*}$ $\rightarrow$ P15$^{*}$ as an example. Fig.~\ref{fig:mahalanobis-histogram} shows the distribution of the Mahalanobis distances for the final Gaussianised chains, while in Fig.~\ref{fig:contours-vs-gauss} we compare these chains to their Gaussian approximations. As can be seen from both of these figures, the chains are well-approximated by normal distributions, thus justifying the use of the Gaussian approximation to compute relative entropies.

\bibliography{main_text_incl_figs}

\end{document}